\documentclass[reprint,notitlepage,showpacs,nobalancelastpage,twocolumn,aps,prb]{revtex4-2}

\usepackage{amsfonts,amssymb,amsmath}
\usepackage{bm}
\usepackage[italicdiff]{physics}
\usepackage{CJK}

\usepackage[english]{babel}
\RequirePackage[T1]{fontenc}

\usepackage{graphicx} 
\usepackage{subfigure}  

\usepackage{siunitx}

\usepackage[hidelinks]{hyperref}
\usepackage{xcolor}

\begin{document}
\begin{CJK*}{UTF8}{gbsn}
\title{Magnetically switchable spin wave retarder with $ 90^\circ$ antiferromagnetic domain wall}
\author{Feiyang Ye  (叶飞扬)}
\affiliation{Center for Joint Quantum Studies and Department of Physics, School of Science, Tianjin University, 92 Weijin Road, Tianjin 300072, China}
\author{Jin Lan (兰金)}
\email[Corresponding author:~]{lanjin@tju.edu.cn}
\affiliation{Center for Joint Quantum Studies and Department of Physics, School of Science, Tianjin University, 92 Weijin Road, Tianjin 300072, China}

\begin{abstract}
Polarization, denoting the precession direction with respect to the background magnetization, is an intrinsic degree of freedom of spin wave.
Using magnetic textures to control the spin wave polarization is fundamental and indispensable toward reprogrammable polarization-based magnonics.
Here we show that due to the intrinsic cubic anisotropy, a $90^\circ$ antiferromagnetic domain wall naturally acts as a spin wave retarder (wave-plate).
Moreover, for a $90^\circ$ domain wall pair developed by introducing a second domain in a homogenous  antiferromagnetic wire, the sign of retarding effect can be flipped by simply switching the magnetization direction of the intermediate domain.
\end{abstract}

\maketitle
\end{CJK*}

\emph{Introduction.}
The core issue of the present information technology is the enormous power dissipation in signal processing.
In conventional electronics, since bits are typically encoded in high/low voltages, heats are unavoidably generated in signal switching due to repeated charging/discharging \cite{Lundstrom210,markov2014limits}.
To ease the dynamical dissipation caused by imbalanced energy flow, it is natural to encode the binary information in spin or polarization, as routinely implemented in
spintronics \cite{wolf2001spintronics,RevModPhys.76.323},
optics \cite{bennett2000quantum,PhysRevLett.118.113901},
acoustics \cite{PhysRevX.8.041027}, as well as magnonics \cite{cheng2016antiferromagnetic,lan2017antiferromagnetic,li2020spin,PhysRevB.93.054412,li2020spin,PhysRevLett.119.056804,lebrun2018tunable} focusing on the spin wave manipulation.

Due to its intrinsic magnetic nature, spin wave can seamlessly interact with magnetic texture, and they together provide an integrated scheme for information harnessing in a purely magnetic way
\cite{lan_SpinWave_2015a,han_Mutual_2019,yu_Magnetic_2020,YU20211}.
With fully unlocked polarization for spin wave in antiferromagnets
\cite{cheng2014spin,wu2016antiferromagnetic,proskurin_SpinWave_2017,zhang2020gate,li2020spin,Ishibashieaaz6931}, their interplay features are remarkably enriched.
The antiferromagnetic skyrmion is shown to induce polarization-dependent magnon Hall effect \cite{daniels2019topological,kim_Tunable_2019}, and a uniaxial $180^\circ$ antiferromagnetic domain wall with Dzyaloshinskii-Moriya interaction (DMI) can induce spin wave polarizing and retarding \cite{lan2017antiferromagnetic}, as well as double refraction \cite{PhysRevB.103.214407}.
Reversely, the motion of domain wall in antiferromagnets can be also controlled via simply tuning the spin wave polarization
\cite{tveten2014antiferromagnetic, yu_2018_polarization,q2018control}.


Despite of the flexibility endowed by polarization from the spin wave side, the development of purely magnetic logic devices is still much impeded by lack of enough reprogrammability from the magnetic texture side.
Indeed, the logic functions of present magnetic devices are mostly realized  by the  trivial existence/absence of magnetic textures \cite{PhysRevLett.93.257202,zhang2015magnetic,PhysRevApplied.12.064053}.
{A promising candidate with intrinsic reprogrammability is the $90^\circ$ domain wall with multiple variants  in materials  with cubic magnetic axes, including $\text{CoFe}_2\text{O}_4$ \cite{bozorth_anisotropy_1955},
$\text{Fe}$ \cite{PhysRevLett.63.1645}, cubic $\text{NiO}$ \cite{baldrati_2019}, Heusler alloys \cite{doi:10.1063/5.0006077}, $\text{Mn}_2\text{Au}$ \cite{vzelezny2014relativistic} and $\text{CuMnAs}$ \cite{wadley2016electrical}}.
The $90^{\circ}$ domain wall has been studied both experimentally and theoretically, including spin wave guiding \cite{lan_SpinWave_2015a} and refraction \cite{hioki2020snell}, and the domain wall motion driven by electric current \cite{tveten2013staggered} or spin-orbit field \cite{gomonay2016high,yang_2019_atomic}.
However, efforts in incorporating both $90^\circ$ domain wall and polarized spin wave to fully unleash the power of magnetic reprogrammability are still lacking.

In this work, we investigate the scattering behaviors of the  polarized spin wave across a  $90^\circ$ antiferromagnetic domain wall.
We show that due to the underlying cubic anisotropy,  a $90^\circ$ antiferromagnetic domain wall shifts the relative phase between  two orthogonal linear spin wave modes.
We further propose that  a $90^\circ$ domain wall pair functions as  a reprogrammable spin wave retarder (wave-plate), where  the sign of retarding effect can be flipped  via switching the central domain direction.
Using cubic anisotropic antiferromagnets for spin wave manipulation extends  current wisdom of designing magnetic devices based on easy-axis or easy-plane magnets
\cite{Parkin190, parkin2015memory, PhysRevLett.121.037202, hanbirefringence, YU20211}.

\begin{figure}[b]
\centering
\subfigure{\includegraphics[width=0.48 \textwidth,trim=60 80 70 60,clip]{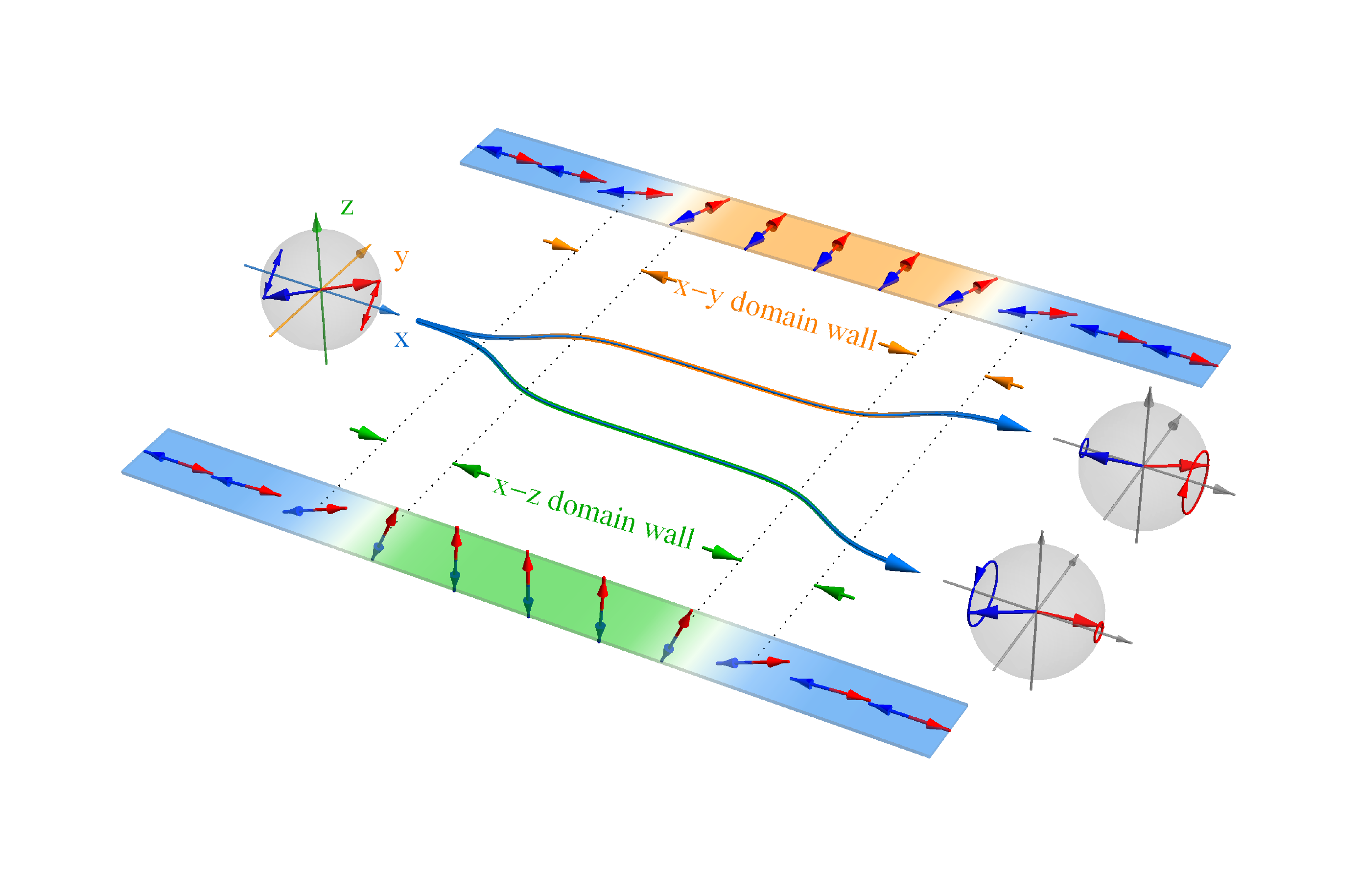}}
\caption{{\bf Schematics of magnetically switchable spin wave retarder (wave-plate) in cubic anisotropic antiferromagnets.}
By preparing a  $y/z$-domain in the center of a homogeneous $x$-domain, a $90^\circ$ antiferromagnetic domain wall pair of $x$-$y$-$x$ (or $x$-$z$-$x$) type forms.
The spin wave retarding effect of such structure can be  magnetically switched:  by changing the intermediate domain to $\hat{\mathbf{y}}/\hat{\mathbf{z}}$ directions, the linear spin wave is transformed to left/right circular spin wave, respectively.
\label{fig:cubic}
}
\end{figure}

\begin{figure*}[t]
\centering
{\includegraphics[width=0.99 \textwidth,trim=0 10 0 0,clip]{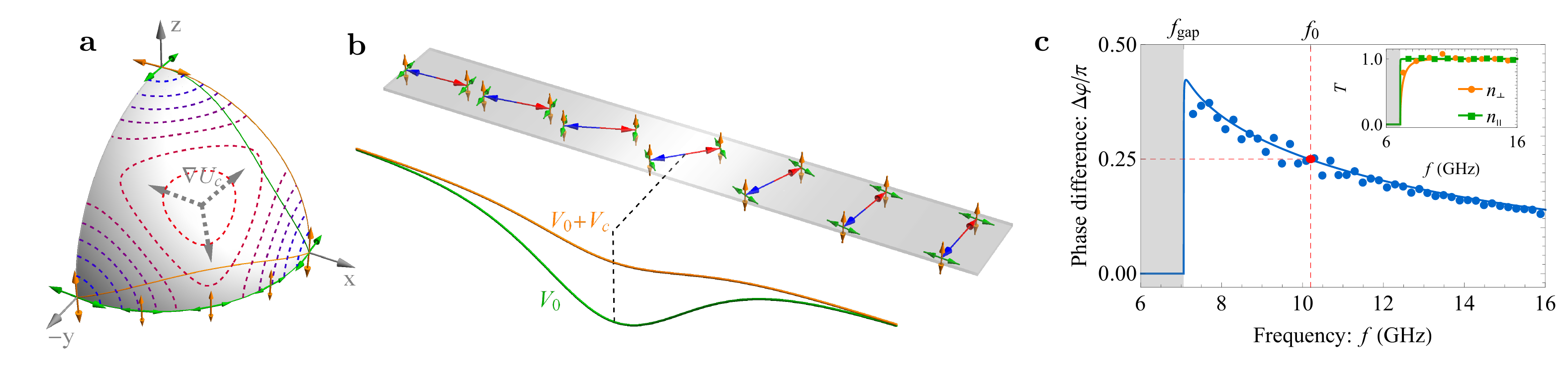}}
\caption{
{\bf The retarding effect of a $90^\circ$ domain wall.}
(a) Contour plots of the cubic anisotropy energy in Bloch sphere.
The green/orange arrows denote the two polarization directions of spin waves, and the green/orange solid lines plot the corresponding total magnetization projected in Bloch sphere.
(b) Schematics of polarized spin waves in a $90^\circ$ domain wall. The red/blue arrows denote the domain wall magnetizations in two sublattices, and the green/orange arrows depict two linearly polarized spin waves. The green/orange lines plot the effective potentials experienced by the in-plane and out-of-plane spin wave modes.
(c) The inter-polarization phase difference $\Delta \varphi$ as a function of spin wave frequency. The solid line plots the full-scattering calculations {\cite{NoteSupp}}, and the dots are extracted from micromagnetic simulations.
The red dot denotes the working frequency $f_0=10.2 ~\si{GHz}$ for the $90^\circ$ domain wall acting as a one-eighth wave-plate.
Inset: the transmission probability $T$ for $n_\parallel$ (green) and $n_\perp$ (orange) respectively.
{The calculations are based on a synthetic antiferromagnet \cite{PhysRevLett.63.1645,duine2018synthetic}} with following magnetic parameters: the inter/intra-sublattice exchange coupling constants $J=1\times 10^6~ \si{A/m}$ and $A= 3.28 \times 10^{-11} ~\si{A}\cdot\si{m}$, the cubic anisotropy constant $K= 3.88 \times 10^4~\si{A/m}$, and the gyromagnetic ratio $\gamma = 2.21\times 10^5~\si{Hz/(A/m)}$.
\label{fig:retarding}
}
\end{figure*}

\emph{Basic model.}
Consider an antiferromagnetic wire with cubic anisotropy lying along $x$-axis as shown in Fig. \ref{fig:cubic}, where the red (blue) arrows denote the magnetization $\mathbf{m}_{1/2}$ in two sublattices respectively.
We define the normalized staggered order $\mathbf{n}=(\mathbf{m}_1-\mathbf{m}_2)/|\mathbf{m}_1-\mathbf{m}_2|$, and the total magnetization $\mathbf{m}=\mathbf{m}_1+\mathbf{m}_2$.
Under the orthogonal constraint $\mathbf{n} \cdot \mathbf{m}=0$, the dynamics of the staggered order $\mathbf{n}$ is governed by the Landau-Lifshitz-Gilbert (LLG) equation for antiferromagnet \cite{tveten2014antiferromagnetic,gomonay2016high,s2016antiferromagnetic,PhysRevB.93.104408,yu_2018_polarization}
  \begin{align}
\label{eqn:LLG_afm}
    \mathbf{n} \times \ddot{\mathbf{n}} = \gamma J \mathbf{n} \times \qty(\gamma \mathbf{h} - \alpha \dot{\mathbf{n}} ),
\end{align}
where $\dot{\mathbf{n}}\equiv \partial_t \mathbf{n}$,  $\ddot{\mathbf{n}}\equiv \partial_t^2 \mathbf{n}$,  $\gamma$ is the gyromagnetic ratio, $\alpha$ is the Gilbert damping constant, and $\mathbf{h}=-(1/\mu_0 M_s)\delta U/\delta \mathbf{n}$ is the effective magnetic field acting upon $\mathbf{n}$, with $\mu_0$ the vacuum permeability and $M_s$ the saturation magnetization.
The magnetic energy $U$ consists of the exchange coupling and cubic anisotropy contributions,
$ U=U_e+U_c= (\mu_0 M_s/2)\int dx \big[ J \mathbf{m}^2+  A (\partial_x \mathbf{n})^2+ K \qty( n_x^2 n_y^2+ n_y^2n_z^2+n_x^2n_z^2) \big] $,
where  $J$  and $A$ are the inter/intra-sublattice exchange constants, and $K>0$ is the cubic anisotropy constant \cite{prabhakar_2009_spin,yang_2019_atomic}.
Due to interweaving sublattices, the long-range dipolar interaction is neglected in this antiferromagnetic environment.

The energy contour of cubic anisotropy in magnetic Bloch sphere is shown in  Fig.~\ref{fig:retarding}(a), where only one-eighth sphere is depicted due to the mirror symmetry with respect to all $3$ Cartesian axes $\hat{\mathbf{x}}_{i}=\qty{\hat{\mathbf{x}},\hat{\mathbf{y}},\hat{\mathbf{z}}}$.
For convenience, we further denote  $i,j,k=\qty{1,2,3}$ as $3$ different indices obeying relation $\hat{\mathbf{x}}_i\times \hat{\mathbf{x}}_j=\varepsilon_{ijk}\hat{\mathbf{x}}_k$ with $\varepsilon_{ijk}$ the Levi-Civita symbol.
As demonstrated, the cubic anisotropic energy $U_c$ minimizes for magnetization $\mathbf{n}$ with one pure Cartesian component $|n_i|=1$ and $n_j=n_k=0$,  increases when the magnetization has mixed components, and maximizes for the magnetization with equal components $|n_i|=|n_j|=|n_k|=1/\sqrt{3}$.
According to above features of cubic anisotropy, there are $3$ different magnetic domains  in $\hat{\mathbf{x}}_i~(i=1,2,3)$ direction as the ground state.
When  $x_i$-domain and $x_j$-domain meet, a $90^\circ$ domain wall arises  with its magnetization lying in the $x_i$-$x_j$ plane, and the out-of-plane magnetic component vanishes  $n_k=0$.

{Under cubic anisotropy, a $180^\circ$ domain wall in between $\pm \hat{\mathbf{x}}_i$ domains is not stable, and naturally splits into two separate $90^\circ$ domain walls, thus is excluded here \cite{PhysRevB.80.094416,doi:10.1063/1.4895803}.}
For simplicity, we call the domains pointing in both $\pm \hat{\mathbf{x}}_i$ directions as the $x_i$-domain, since their difference in polarization manipulation is minor here.

\emph{Polarization-dependent spin wave scattering.}
When spin wave is excited within a uniform domain, its two linear polarization modes are degenerate since the cubic anisotropy is invariant under the interchange of any two axes, as shown in Fig.~\ref{fig:retarding}(a).
As spin wave propagates upon the domain wall, these two linear polarizations become  in-plane and  out-of-plane  modes  in reference to  the domain wall plane, as illustrated in Fig. \ref{fig:retarding}(b).
{For in-plane modes, magnetization oscillations upon domain wall are still with two magnetization components, thus the spin wave dynamics is almost unaffected by the cubic anisotropy; while the out-of-plane modes involve the $3$rd magnetization component, thus are subject to additional suppression of the cubic anisotropy.}
Due to the distinct behaviors of in-plane and out-of-plane modes under the cubic anisotropy, polarization-dependent spin wave scattering thus arises in a $90^\circ$ domain wall.

To proceed the investigations on  the spin wave scattering, we denote the $90^\circ$ domain wall as $\mathbf{n}_0$, and the spin wave as $\delta \mathbf{n}=n_\parallel \hat{\mathbf{e}}_\parallel + n_\perp \hat{\mathbf{e}}_\perp$, where $ \hat{\mathbf{e}}_\parallel$ and $\hat{\mathbf{e}}_\perp$ are the in-plane and out-of-plane directions with respect to the domain wall plane subtended by $\mathbf{n}_0$.
For $x$-$y$ domain wall, its magnetic profile is $\mathbf{n}_0=\qty(\sqrt{[1-\tanh(x/W)]/2}, \sqrt{[1+\tanh(x/W)]/2},0)$, with $W=\sqrt{A/K}$ the characteristic width, and the $x$-$z$ ($y$-$z$) domain walls take similar form \cite{tveten2013staggered}.
With the above domain wall profile, the spin wave dynamics is then recast from LLG equation \eqref{eqn:LLG_afm} to Klein-Gordon-like equations
\begin{subequations}
\label{eqn:sw_eom}
\begin{align}
        -\ddot{n}_\parallel  = & \gamma^2 J \left[-A\partial_x^2 + V_0(x) \right] {n}_\parallel,\\
        -\ddot{n}_\perp =& \gamma^2 J \left[-A\partial_x^2 + V_0(x)+V_c(x)\right] {n}_\perp,
\end{align}
\end{subequations}
where the potential $V_0= K[1-2\sech^{2}(x/W)]$ is caused by the inhomogeneous domain wall profile, and $V_c = (5/4)K\sech^{2}(x/W)$ solely emerging for the out-of-plane mode is induced by the cubic anisotropy.
The potential $V_0$ is the celebrated P{\"o}schl-Teller type potential sharing the same profile as the $180^\circ$ domain wall case \cite{yan2011all,kim2014propulsion,tveten2014antiferromagnetic,yu_2018_polarization}, which always passes  spin wave perfectly.

From Eq.~(\ref{eqn:sw_eom}), in homogeneous domains far away from the domain wall ($|x|\gg W$), the in-plane and out-of-plane modes are degenerate with dispersion relation $\omega = \gamma \sqrt{J(Ak^2+K)}$, with $\omega$ the spin wave frequency  and $k$ the wavevector.
Inside the domain wall, the degeneracy of two modes is lifted by the potential $V_c$, leading to a phase difference accumulated between two modes across the domain wall.
Fig.~\ref{fig:retarding}(c) plots the  inter-polarization phase difference $\Delta \varphi\equiv \varphi_{\perp}-\varphi_{\parallel}$ obtained from full scattering calculations {\cite{NoteSupp,PhysRevB.72.035450}} and extracted from micromagnetic simulations, which overlap well  with each other.
The positive phase difference $\Delta\varphi>0$, together with almost perfect transmission of both two modes (except some reflections in the low frequency range) as demonstrated in Fig. \ref{fig:retarding}(c), corroborate  the retarding effect of the $90^\circ$ domain wall.

For $x_i$-$x_j$ domain wall, the out-of-plane direction is always $\hat{\mathbf{e}}_\perp=\hat{\mathbf{x}}_k$, while the in-plane direction gradually changes $\hat{\mathbf{e}}_\parallel = \hat{\mathbf{x}}_j \to \hat{\mathbf{x}}_i$, indicating a rotation of polarization basis.
In addition, for $x_i$-$x_j$ and $x_i$-$x_k$ domain walls starting from  the same $x_i$-domain, the roles of linear-$x_j$ (linear-$x_k$) modes as  in-plane (out-of-plane) mode exchange, as illustrated in Fig. \ref{fig:retarding}(a) {\cite{NoteSupp}}.
Moreover, when linear-$x_j$ (linear-$x_k$) mode penetrates through $3$  cyclically connected $x_i$-$x_j$-$x_k$-$x_i$ domain walls, it is transformed into the orthogonal linear-$x_k$ (linear-$x_j$) mode, rather than restoring its original polarization direction.
These unique polarization basis variation features in cubic anisotropic system highlight nonabelian behaviors of  $\mathrm{SU}(2)$ spin rotation \cite{RevModPhys.82.1959}.

\begin{figure}[t]
\centering
{\includegraphics[width=0.48 \textwidth ,trim=5 10 5 5,clip]{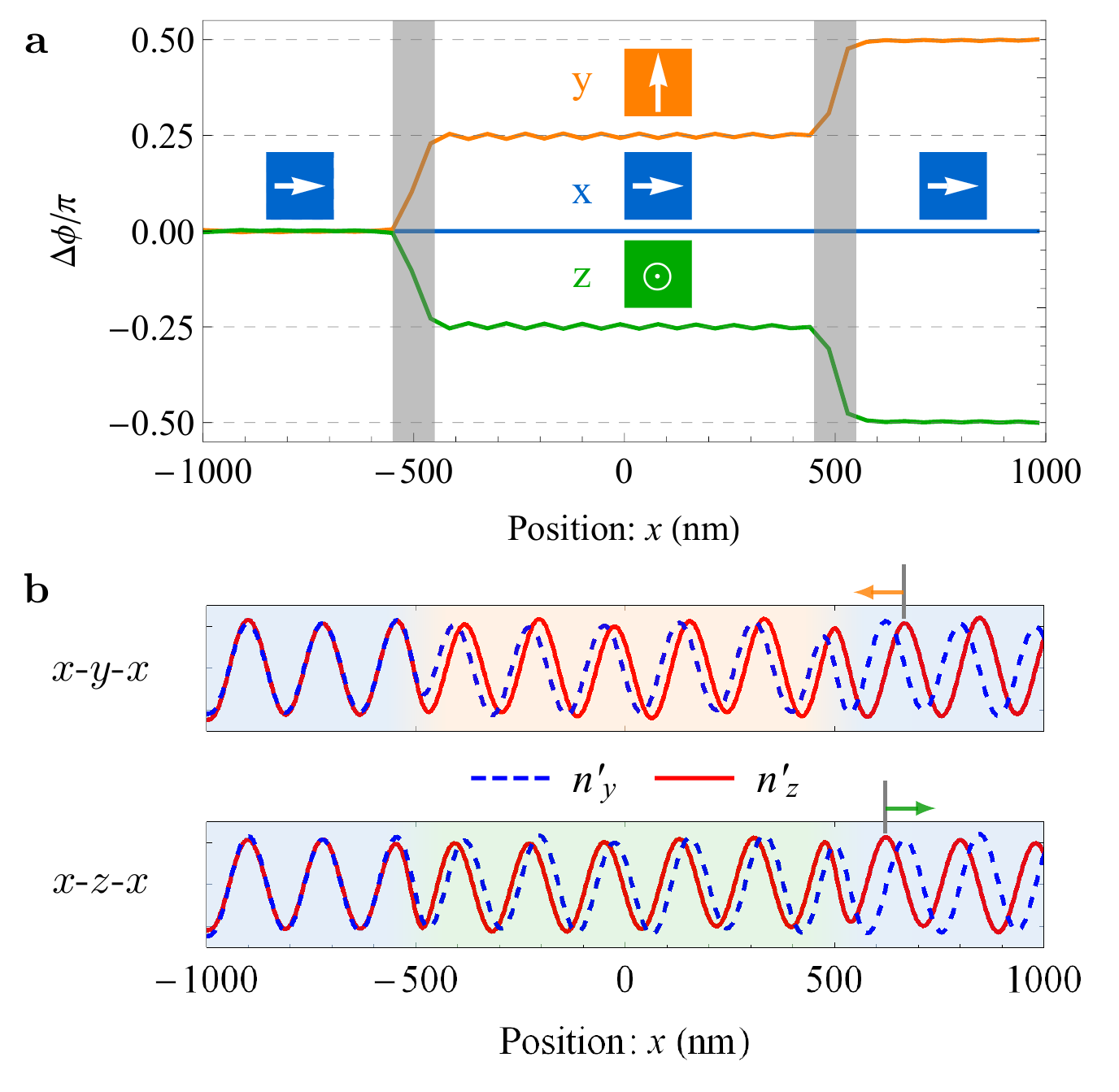}}
\caption{
{\bf Micromagnetic simulations of the spin wave retarder based on a domain wall pair.}
(a) The spatial evolution of inter-polarization phase difference $\Delta \phi$ extracted from micromagnetic simulations. The orange/green lines are for the $x$-$y$($z$)-$x$  domain wall pair, and the blue line is for homogenous $x$-domain in reference.
(b) The spatial profile of the spin wave components in the $x$-$y$-$x$ (upper panel) and $x$-$z$-$x$ (lower panel) domain wall pair.
The blue/red lines plot the ${n}'_y$ and ${n}'_z$ components, which refers to $n_{y/z}$ in $x$-domain and their extensions within domain wall pair.
In (a) and (b), two domain walls are located at $x=\pm 500~\si{nm}$, the spin wave frequency is $f=10.2~\si{GHz}$, the magnetic parameters follow Fig. \ref{fig:retarding},  and the damping constant is artificially set to $\alpha=0$ for better demonstration of the retarding effect.
\label{fig:retarder}
}
\end{figure}

\emph{Spin wave retarder.}
The rotation of polarization basis together with the phase difference $\Delta\varphi$ indicate that  a  $90^\circ$ domain wall acts as a rotated retarder \cite{goldstein2017polarized} for spin wave.
For pure retarding functionality,  a pair of $90^\circ$ domain walls are necessary,  where the basis rotation is cancelled, while the phase difference is doubled.
Such a domain wall pair can be straightforwardly prepared by placing a second domain in a homogeneous magnetic wire, as illustrated in Fig. \ref{fig:cubic}.
In between two $x_i$-domains, the intermediate domain may either be $x_j$-domain or $x_k$-domain to form such a domain wall pair.
And for these two configurations, the in-plane and out-of-plane directions exchange, leading to opposite sign of the phase difference  between linear-$x_j$ and linear-$x_k$ components, i.e. $\Delta\phi \equiv \varphi_k-\varphi_j=\pm 2\Delta\varphi$ (see details in Fig.~S1).


The retarding effect of the domain wall pair is verified by micromagnetic simulations, as plotted in Fig.~\ref{fig:retarder}.
A $45^\circ$-linear spin wave with working frequency $f_0=10.2~\si{GHz}$ is injected from the left side of an $x$-$y$-$x$ (or $x$-$z$-$x$) domain wall pair.
The phase difference $\Delta \phi$ levels off inside each domain, but increases/decreases by $\pi/4$ as passing through each $x$-$y$ (or $x$-$z$) domain wall in Fig.~\ref{fig:retarder}(a).
After the spin wave reaches the $x$-domain on the right side, it acquires a phase difference of $\pm \pi/2$ as expected, indicating that the $45^\circ$-linear spin wave becomes left/right circularly polarized, respectively.
Typical snapshots of spin wave propagating in the domain wall pair are further depicted in Fig.~\ref{fig:retarder}(b).
For convenience, we denote ${n}'_{y/z}$ as extensions of the  spin wave components $n_{y/z}$ in the $x$-domain to the whole domain wall pair region.
While overlapping on the left side, the ${n}'_y$ component lags/advances the ${n}'_z$ component as penetrating through the $x$-$y$-$x$ (or $x$-$z$-$x$) domain wall pair,
highlighting their  opposite retarding effects.

\begin{figure}[t]
\centering
{\includegraphics[width=0.48 \textwidth, trim=10 5 10 5, clip]{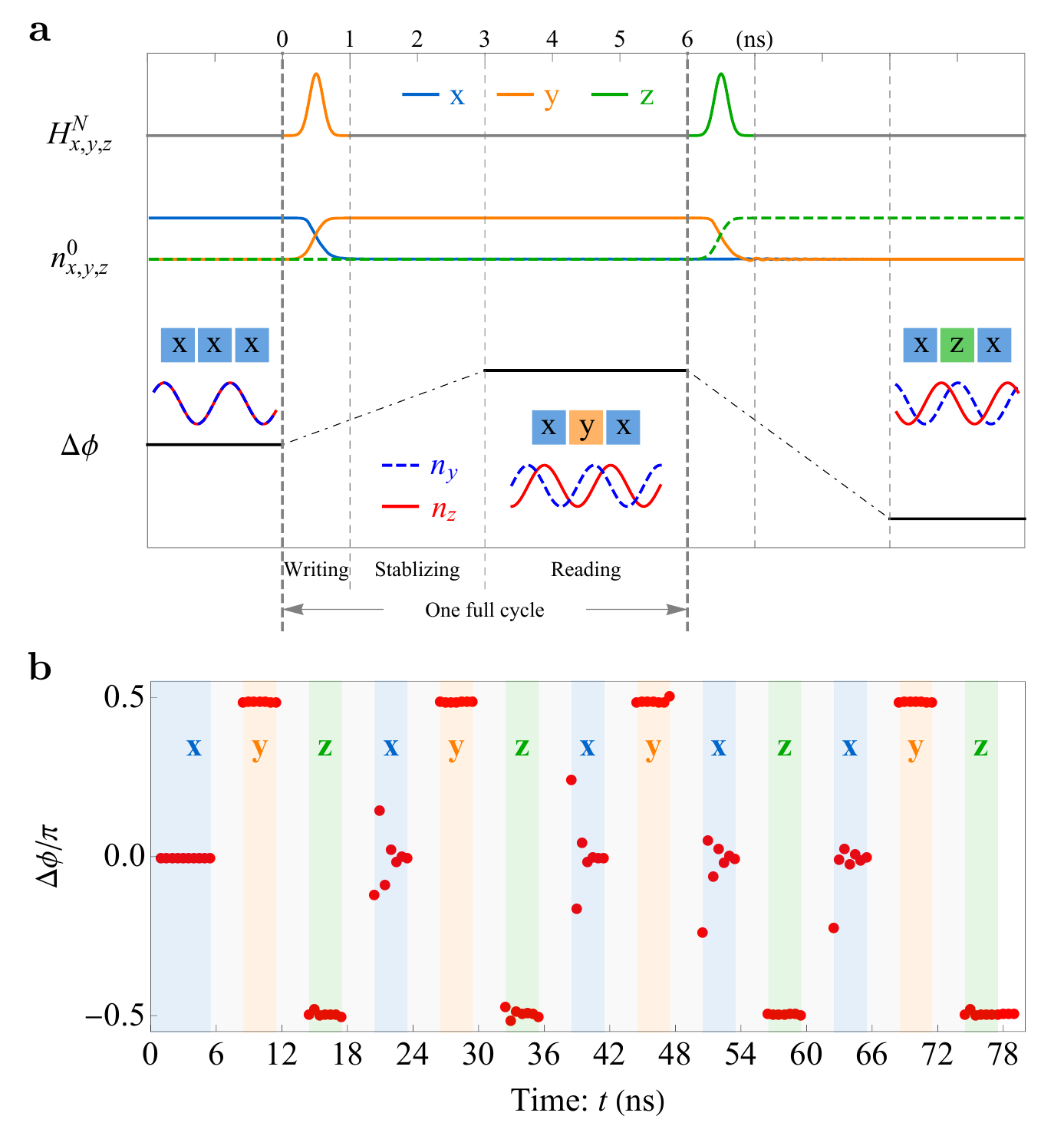}}
\caption{
{\bf Reprogrammable spin wave retarder by switching the magnetization of the central domain. }
(a) Timing diagram of magnetic signals in one programming cycle of $6~\si{ns}$.
A N{\'e}el-order magnetic field $\mathbf{H}^N$ is applied at $[0,1]~\si{ns}$ for central domain writing, the magnetic structure $\mathbf{n}_0$ waits for stabilization at $[1,3]~\si{ns}$, while the inter-polarization phase difference $\Delta\phi$ is read at $[3,6]~\si{ns}$.
The insets schematically show the state of the domain wall pair, and the spatial profile of two spin wave components at the right $x$-domain.
(b) Time evolution of phase difference $\Delta\phi$ in $12$ programming cycles.
The red dots are phase difference extracted from the simulations, and the background blue/orange/green colors are for the stabilized $x/y/z$-domain respectively.
In (a) and (b), the magnetic wire lies at $[-1000,1000]~\si{nm}$, the domain walls are prepared at $\pm 350~\si{nm}$,  the $45^\circ$-linear spin wave with frequency $f=10.2~\si{GHz}$ is continuously excited at $x=-500~\si{nm}$, and output spin wave is read at $x=600~\si{nm}$.
The Gilbert damping constant inside the wire is $\alpha=1.4 \times 10^{-2} $, and is increased to $\alpha=0.1$ near the boundaries ($|x|>900~\si{nm}$) to absorb redundant spin waves.
\label{fig:reprogram}
}
\end{figure}

\emph{Magnetic reprogrammability.}
{Antiferromagnetic domain can be switched to different directions by exerting an effective N{\'e}el-order magnetic field induced by current pulses
\cite{baldrati_2019,vzelezny2014relativistic,wadley2016electrical,cheng2020electrical,bodnar2018writing}.}
By rewriting the central domain using such staggered field, the retarding effect of the domain wall pair is then reprogrammed, as shown in Fig.~\ref{fig:reprogram}.
When a N{\'e}el-order magnetic field $\mathbf{H}^N=H^N_y\hat{\mathbf{y}}$ of duration $1~\si{ns}$ is applied, the central domain quickly switches from  $\hat{\mathbf{x}}$ direction to the designated $\mathbf{n}_0=\hat{\mathbf{y}}$ direction after a stabilization time of $2~\si{ns}$, and  the spin wave phase difference $\Delta\phi$ quickly changes from $0$ to $0.5\pi$ in Fig. \ref{fig:reprogram}(a).
The signals in the last $3~\si{ns}$ of the $6~\si{ns}$ full cycle become  stable and clean, thus the retarder is ready for the next programming cycle.

In Fig.~\ref{fig:reprogram}(b), a train of $12$ external field pulses are applied to switch the central domain arbitrarily between  $3$ orthogonal directions {\cite{NoteSupp}}.
For all $12$ cycles, the output phase difference $\Delta\phi$ steadily switches to  $0$ or $\pm \pi/2$,  when the central domain is stabilized in $\hat{\mathbf{x}}/\hat{\mathbf{y}}/\hat{\mathbf{z}}$ directions.
The switching to central $x$-domain corresponds to erasion of the domain wall pair, thus abundant spin wave is released, leading to slightly lower signal quality.
Despite of fluctuations, the phase difference $\Delta\phi$ still  clearly falls into three well separated ranges, highlighting the robustness of the reprogrammable retarder.

\emph{Discussion.}
The retarding effect also arises in $180^\circ$ domain wall with easy-axis anisotropy in Ref.~\onlinecite{lan2017antiferromagnetic}, but the underlying mechanisms are distinct:
the degeneracy of polarizations is lifted by the intrinsic cubic anisotropy in this work, but relies on the external DMI in Ref.~\onlinecite{lan2017antiferromagnetic}.
In addition, the cubic anisotropy preserves high symmetry, while the DMI reversely requires to  break the inversion symmetry.
The involvement of only exchange coupling and cubic anisotropy also implies that the retarding effect in this work is universally applicable to all cubic anisotropic antiferromagnets.


To switch the domain wall pair between two configurations (e.g. $x$-$y$-$x$ and $x$-$z$-$x$ configurations), one simply needs to change the  magnetic state of the central domain.
These two configurations are symmetric with each other, sharing similar structure and exactly the same energy, but introducing the opposite retarding effect.
In this sense, the reprogrammability demonstrated here is truly symmetric, and distinguishes from the previous proposals  simply based on the presence/absence states.
To guarantee the symmetry of the  reprogrammability, triplet degeneracy is required, with one state as the reference and the other two for switching, e.g. the triple equivalent domains hosted by the cubic anisotropy.
The spirit of symmetric reprogrammability thus coincides with the polarization (spin)-based manipulations via exploring the additional degree of freedom, rather than breaking symmetry in existing degrees of freedom.

\emph{Toward experiments.}
The polarized spin wave can be excited by applying an oscillating magnetic field via an antenna, and then detected via the antiferromagnetic spin pumping by attaching an adjacent $\text{Pt}$ layer \cite{cheng2014spin,vaidya2020subterahertz}, which converts the left/right circular polarizations to positive/negative electrical signals.
In addition, the application of a global N\'eel-order magnetic field should remove remanent multi-domains in the wire.

\emph{Conclusion.}
In conclusion, we demonstrate that a $90^\circ$ antiferromagnetic domain wall pair naturally functions as a reprogrammable spin wave retarder.
By switching the magnetic direction of the central domain, the linear spin wave can be converted to either left- or right- circular mode.
Due to the rich functionality and embedded reprogrammability, the cubic anisotropic antiferromagnet manifests as a new platform for magnetic information processing.

\emph{Acknowledgments.}
J.L. is  grateful to Jiang Xiao for insightful discussions.
This work is supported by  National Natural Science Foundation of China (No. 11904260), and the Startup Fund of Tianjin University.


%

\end{document}